\documentclass[a4paper,11pt]{article}
\pdfoutput=1 

\usepackage{jcappub} 

\usepackage[T1]{fontenc} 
\usepackage{comment}
\usepackage{xspace}

\def\qgs{QGSJ\textsc{et}\,-II.04\xspace}

\title{\boldmath Gamma/hadron discrimination at high energies through the azimuthal fluctuations of air shower particle distributions at the ground}

\author[a,b]{R. Concei\c{c}\~ao}
\author[a,b]{L. Gibilisco}
\author[a,b]{M. Pimenta}
\author[a,b]{B. Tom\'e}

\affiliation[a]{Laboratório de Instrumentação e Física Experimental de Partículas (LIP),\\ Av. Prof. Gama Pinto, 2, P-1649-003 Lisbon, Portugal}
\affiliation[b]{Departamento de F\'isica, Instituto Superior T\'{e}cnico (IST),\\ Universidade de Lisboa, Av. Rovisco Pais 1, 1049-001, Lisbon, Portugal}

\abstract{Wide field-of-view gamma-ray observatories must fight the overwhelming cosmic ray background to identify very-high-energy astrophysical gamma-ray events.
This work introduces a novel gamma/hadron discriminating variable, $LCm$, which quantifies the azimuthal non-uniformity of the particle distributions at the ground. This non-uniformity, due to the presence of hadronic sub-showers, is higher in proton-induced showers than in gamma showers. The discrimination power of this new variable is then discussed, as a function of the air shower array fill factor, in the energy range $10\,$TeV to $1\,$PeV, and compared to the classical gamma/hadron discriminator based on the measurement of the number of muons at the ground. The results obtained are extremely encouraging, paving the way for the use of the proposed quantity in present and future large ground-array gamma-ray observatories.}

\begin{document}
\maketitle
\flushbottom

\section{Introduction}
\label{sec:introd}

The recent detection of gamma rays with energies up to the PeV~\cite{Peta_gamma,LHAASO_PeV} has opened a new window to the extreme energy Universe. Due to the low fluxes and the high background charged cosmic ray, such detection is only possible at ground-based gamma-ray observatories with large surface areas (of the order or higher than $ 1\,{\rm km^2}$) and able to discriminate, with very high efficiency, hadron showers from gamma showers. 

At high energies (above tens of TeV),  high background rejection factors may be reached by studying the distribution of the particles at the ground as a function of the distance to the shower core, identifying the existence of energetic sub-showers, or measuring the steepness, compactness or bumpiness of the Lateral Distribution Function (LDF) \cite{Greisen,Abeysekara_2017,LATTES}, or, in the case of Cherenkov telescopes, the differences on the shower longitudinal development~\cite{Hillas,Weekes}. On the other hand, the measurement of the number of muons arriving at the ground \cite{Blake_1995,Hayashida_1995,APEL2010202} is, whenever possible, one of the best discriminators. The measurement of muons can be done by absorbing the electromagnetic component of the shower by shielding the detectors using earth (e.g. \cite{1994NIMPA.346..329B,Aab_2021,LHAASO_muon}), water  (e.g. \cite{MARTA}, \cite{DLWCD_Antoine}), concrete or some other inert material. Alternatively, muons might be detected by studying, in detectors with several light sensors, the differences in time and/or intensity of the collected signals (e.g. \cite{zuniga2017detection,Gonzalez_2020, Mercedes}).
In the end, a global rejection factor of the order or higher than $10^4$ should be achieved.

In this article, we explore the azimuthal non-uniformity of the particle distributions at the ground, introducing a new variable, $C_{k}$. This quantity is computed in successive circular rings centred at the shower core with a radius $r_{k}$ in the shower transverse plane. From the $C_{k}$ distribution as a function of $r_k$ is then defined a new gamma/hadron discriminating variable, $LCm$, being simply the value of the $\log(C_{k})$ distribution at a given $r_k$.
The paper is organised as follows: the simulation sets used to evaluate all the findings presented in this paper are described in section~\ref{sec:simulation}; the variable $C_{k}$  is introduced in section \ref{sec:Ck}, while the discriminator quantity $LCm$ is presented and discussed in section \ref{sec:LCm}. There, the discrimination power of $LCm$ is tested for different array fill factors $(FF)$ in the energy range from $10\,$TeV to $1\,$PeV; 
in section \ref{sec:Ck-MU} the correlation of this new variable with the logarithm of the number of muons at the ground is analysed; finally, in section  \ref{sec:conclusions} the use of this new variable in the present and future large ground array gamma-ray observatories is discussed.

\section{Simulation sets}
\label{sec:simulation}
CORSIKA (version 7.5600) \cite{CORSIKA} was used to simulate gamma-ray and proton-induced vertical showers assuming an observatory altitude of $5200\,$m a.s.l. The shower energy ranged from $10\,$TeV up to $2\,$PeV, being generated with an $E^{-1}$ energy spectrum. FLUKA~\cite{fluka,fluka2} and \qgs \cite{qgs} were used as hadronic interaction models for low and high energy interactions, respectively.

A ground detector array was emulated by a 2D-histogram with cells with an area of $\sim 12\,{\rm m^2}$ covering all the available ground surface with a fill factor equal to one (FF$=$1). Fill factors smaller than one were obtained by masking the 2D-histogram with regular patterns. 

Each cell represents a station. The signal in each station was estimated as the sum of the expected signals due to the particles hitting the station, using dedicated parameterizations as a function of the particle energy for protons, muons and electrons/gammas. These curves were obtained by injecting vertical particles sampled uniformly on top of a small water Cherenkov detector station with four PMTs placed at the bottom~\cite{Borja4PMTs}. The parameterizations were built for the mean signal in the station and the signal distribution standard deviation. Through the use of these two numbers, it is then possible to emulate the WCD signal response fluctuations due to the stochastic processes of particle interactions and light collection. Additionally, for muons, it was also included the fluctuation in their tracklength due to geometry variations. Such was achieved using the distribution of the muon taken from proton-induced shower simulations ran over a Geant4 simulation, which provided the geometry of the WCD array and stations.

To make the comparisons fair and mimic realistic experimental conditions, the energy reconstruction is emulated conservatively by taking the total electromagnetic signal at the ground for each event.
An energy bin with size $\log(E/{\rm GeV}) = 0.2$ is taken for gammas, and cuts on the electromagnetic signal are derived. These cuts were defined taking $\mu \pm \sigma$, where $\mu$ is the mean of the distribution of the sum of the electromagnetic signal collected in the stations 
and $\sigma$ is its standard deviation.

These cuts are applied to the proton simulations over an extended energy range. In this way, we are comparing shower events with the same total signal at the ground.

\section{The $C_{k}$ variable}
\label{sec:Ck}

The new variable $C_{k}$ is defined for each radial ring $k$ as:

\begin{equation}
C_{k} =\frac{2}{n_{k}(n_{k}-1)} 
\frac{1}{\left<S_{k}\right>}\sum_{i=1}^{n_{k}-1}\sum_{j=i+1}^{n_{k}}(S_{ik}-S_{jk})^{2} ,   
\label{eq:CK}
\end{equation}

where $n_{k}$ is the number of stations in ring $k$, $\left<S_{k}\right>$ is the mean signal in the stations of the ring $k$, and $S_{ik}$ and $S_{jk}$ are the collected signals in the stations  $i$ and $j$ of the ring $k$, respectively. The term $\frac{2}{n_{k}(n_{k}-1)}$ is the inverse of the number of two-combinations for $n_k$ stations, $\binom{n_k}{2}$. In this work, each circular ring $k$ is centred around the shower core position and has a width of $10$\,m.

The variable $C_{k}$ is just the mean sum of the square of the differences between the collected signals in any pair of two stations in the ring $k$ normalised to the mean signal observed in one station of the ring $k$. This normalisation should minimise the possible correlations between the dependence in $r_{k}$ of $C_{k}$ and the lateral distribution function. On a side note, the information in the LDF could be combined with the azimuthal asymmetry information, $C_k$, to enhance the discrimination capability. This is out of the scope of this paper and will be addressed in a future work.

For a completely uniform azimuthal distribution,  $C_{k}$ is, by construction, equal to zero, while it should be greater than zero otherwise. Due to the presence of hadronic sub-showers, the pattern at the ground of a proton shower is more complex than the pattern of a gamma shower, whose development is basically driven by electromagnetic interactions. Thus, $C_{k}$ is expected to be higher in the case of proton-induced showers compared to gamma showers of equivalent energies at the ground. 

\begin{figure}[tbp]
 \centering
\includegraphics[width=0.8\textwidth]{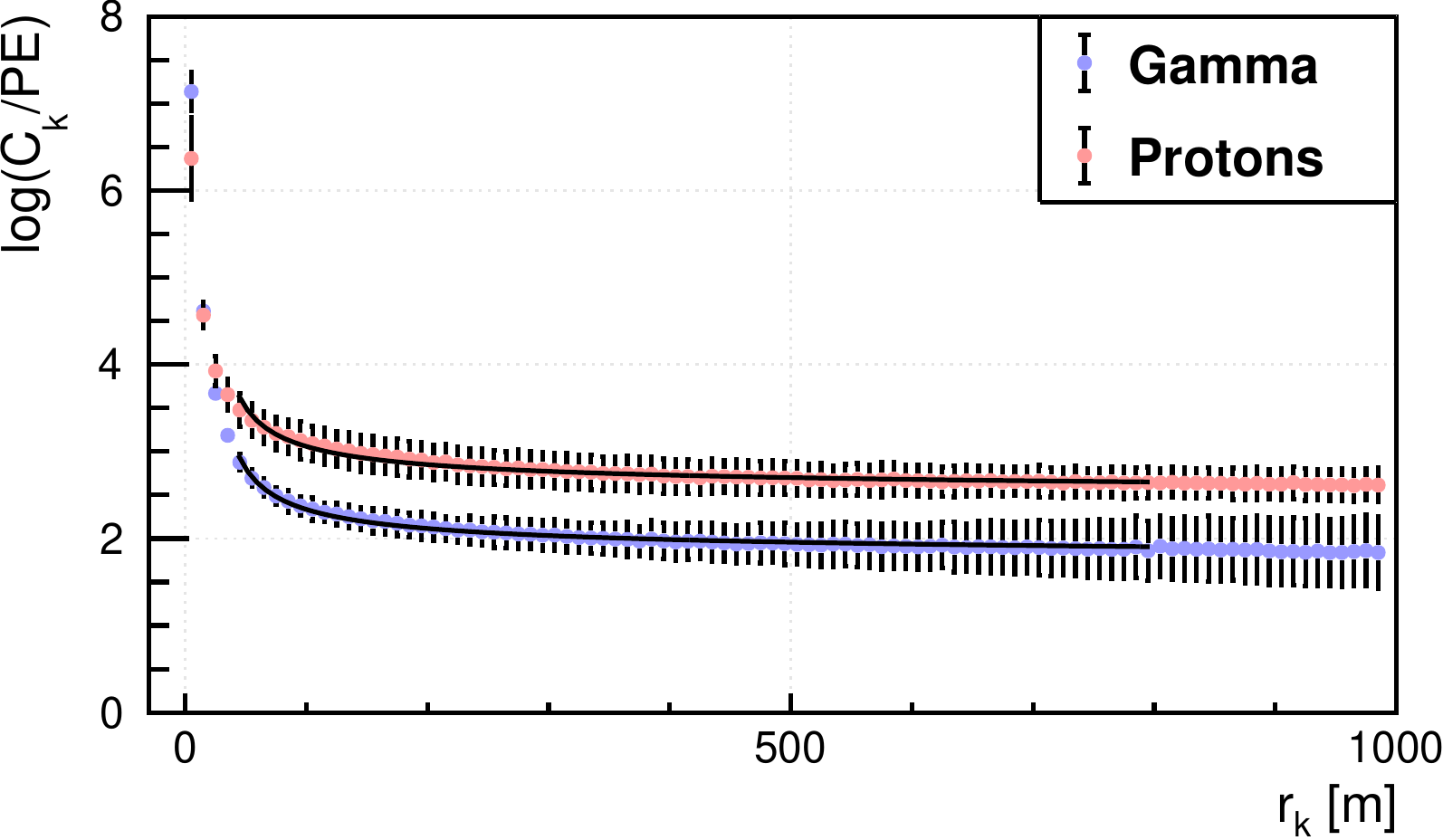}
\caption{\label{fig:CK_100TeV_FF1}  Mean value of log($C_{k}$)  as a function of $r_{k}$ for gamma showers with a primary energy $\in [100; 160]\,$TeV  (blue, lower points) and for proton showers (red, upper points) with  similar energies at the ground considering an array $FF=1$. The errors bars are the RMS of the log($C_{k})$ distributions in each $r_{k}$ bin. The full lines represents the best fit using the parametrisation  expressed in equation \ref{eq:CKfit}.
 }
\end{figure}

In fact, this behaviour is clearly seen for $r_{k}$ greater than $40\,$m in figure \ref{fig:CK_100TeV_FF1}, where the distributions of the mean values of the log($C_{k}$) variable are represented as a function of $r_{k}$ for gamma (blue points) and proton (red points) showers with $\sim100\,$TeV considering a $FF=1$. The energy of the gamma-induced showers are in the energy interval of $100\,$TeV to $160\,$TeV, while proton-induced showers have similar energies at the ground. 

The $C_{k}$ distributions as a function of $r_k$ depend on the primary energy and the array fill factor, but, if $C_{k}$ would be somehow correlated to the average number of detected muons at ground (see section \ref{sec:Ck-MU}), they should scale with a $K$ factor defined as:

\begin{equation}
 K = E^{\beta} \times  FF  ,
 \label{eq:ruleThumb}
\end{equation}

where $E$ is the primary gamma-ray energy (in TeV), $\beta$  is the index of the power dependence of the mean number of muons at the ground, and $FF$ the fill factor defined in the interval $]0;1]$. The parameter $\beta$ was fixed to $0.925$, which is a typical mean value used in hadronic shower simulations \cite{CONEX}.

\begin{figure}[tbp]
 \centering
\includegraphics[width=0.8\textwidth]{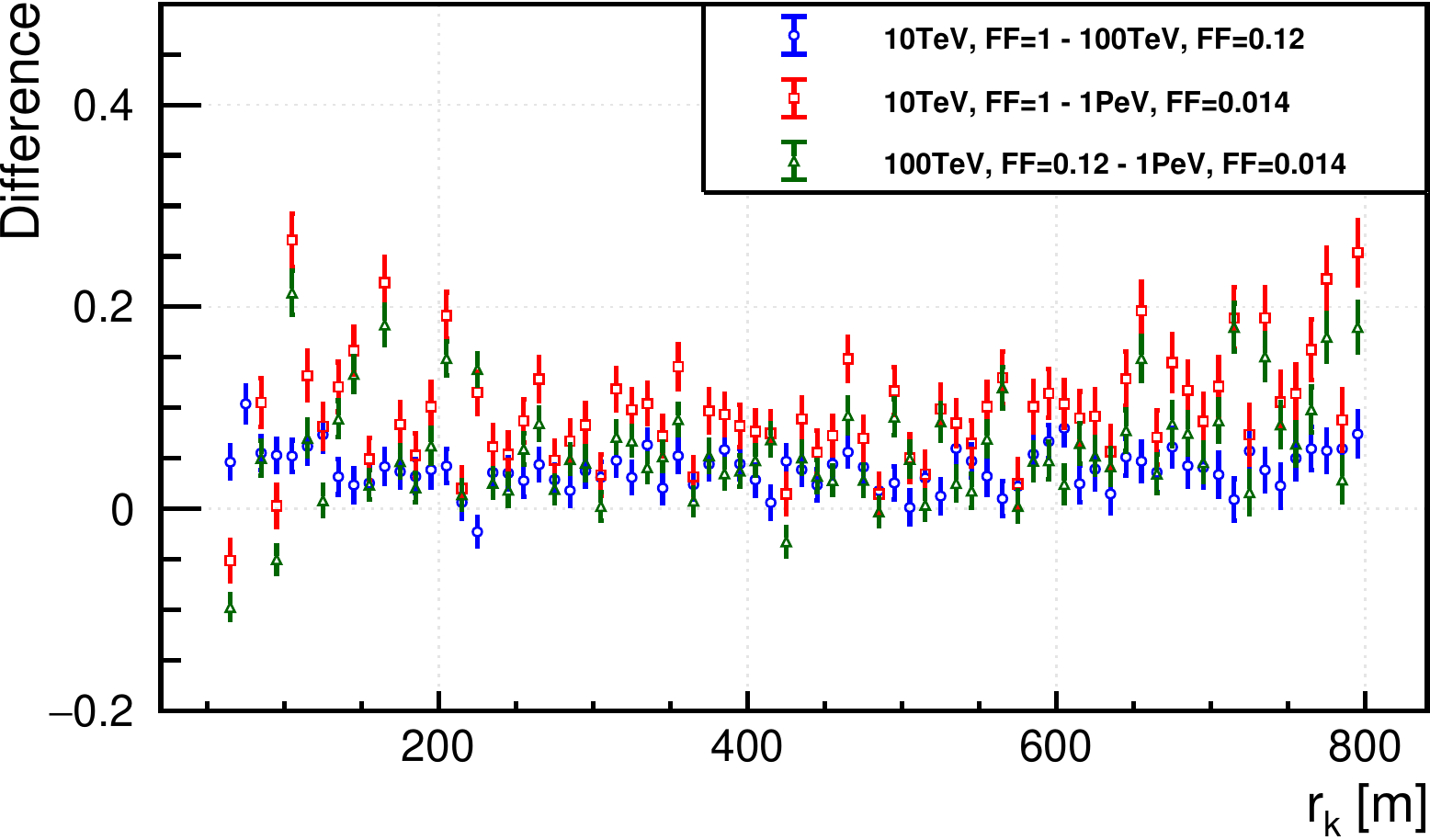}
\caption{\label{fig:rule_of_thumb}   Difference of the average distributions of $\log(C_k)$ as a function of $r_{k}$ of two simulation sets with different energies $E$ and fill factors $FF$ (see legend for details) but with  similar $K \sim 10.6$ factors, computed using equation \ref{eq:ruleThumb}. This plot was produced using vertical proton-induced showers. The displayed error bars represent the propagated statistical uncertainty.
 }
\end{figure}

Hence, for identical $K$ factors but different energies and fill factors, the mean values of the log($C_{k}$) distributions is expected to be essentially identical. This crude rule of thumb is successfully verified, for $r_{k} > 100\,$m, at a level of a few percent in figure \ref{fig:rule_of_thumb}. In this figure are shown the difference of the average distributions of $\log(C_{k})$ as a function of $r_k$ such that the compared showers have similar $K$ factors, namely: $E \approx 13\,$TeV\footnote{the mean value of the chosen bin assuming an $E^{-1}$ spectrum as a balance between computational time and statistics at the highest energies.}, $FF=1$ $\rightarrow$ $K = 10.67$;  $E \approx 130\,$TeV, $FF=0.12$ $\rightarrow$ $K = 10.77$; and $E \approx 1.3 \,$PeV, $FF=0.014$ $\rightarrow$ $K = 10.58$. 

Finally, it should be also noted that the global behaviour of the $\log(C_{k})$ distributions in the region $r_{k} > 40$m can  be described  by the following parametrisation 
(full lines in figure \ref{fig:CK_100TeV_FF1}): 

\begin{equation}
\log(C_{k}) = a  +  \frac{b}{\log \left(\frac{r_{k}}{40\,{\rm m}}\right)+1} 
\label{eq:CKfit}
\end{equation}

where $a$ and $b$ are free real parameters.
In the limit $ r_{k}\rightarrow 40 $ m, $\log(C_{k40}) = a+b$ and in the limit $ r_{k}\rightarrow{\infty}$, $\log(C_{k \rightarrow \infty}) \rightarrow a$.

This simple parametrisation complies with the two asymptotic limits and indeed captures well the evolution of  $C_{k}$ as a function of  $ r_{k}$. The parameters $a$ and $b$ are determined, for each event, by fitting the corresponding $\log(C_{k})$  distributions to the above parametrisation, using as errors the RMS of the $\log(C_{k})$ distributions in each $r_{k}$ bin, which are also displayed in figure \ref{fig:CK_100TeV_FF1}.

The overall quality of the fit is good ( $ \chi^2 / n.d.f. \sim 1$ ) with a very small tail of events $(<1\%)$ with $ \chi^2 / n.d.f.  > 2 $. The analysis of these tail events is out of the scope of the present article, so these events were simply discarded. However, this type of analysis may tag events where extreme fluctuations in the shower development had occurred, such as, for instance, the so-called double-bang events or, more speculatively, exotic physics in the air shower development.


\section{ The gamma/hadron discrimination variable: $LCm$}
\label{sec:LCm}

A simple new  $\gamma / {\rm h}$ discriminator is defined  as the value of the $\log(C_{k}) $ at a given value of $ r_{k} = r_{m} $, computed using the  parametrisation introduced in equation \ref{eq:CKfit}: 

\begin{equation}
LCm \equiv \left. \log\left(  C_{k}  \right) \right|_{r_k = r_m} , 
\end{equation}

The value of $r_{m} $ could be optimised for each energy bin as the value corresponding to the largest separation between the gamma and proton distributions.
 


The optimal $r_{m}$ value should grow slowly in a monotonous way as a function of the shower energy $E$. Nevertheless, to allow a more straightforward comparison between the values of $LCm$ at the different energies in the interval between $10\,$TeV and $1\,$PeV, and considering that the sparse array  may have a radius of many hundreds of meters,  $r_{m}$ was fixed to be:

\begin{equation}
r_{m} = 360 \, m   . 
\end{equation}

\begin{figure}[tbp]
 \centering
\includegraphics[width=0.8\linewidth]{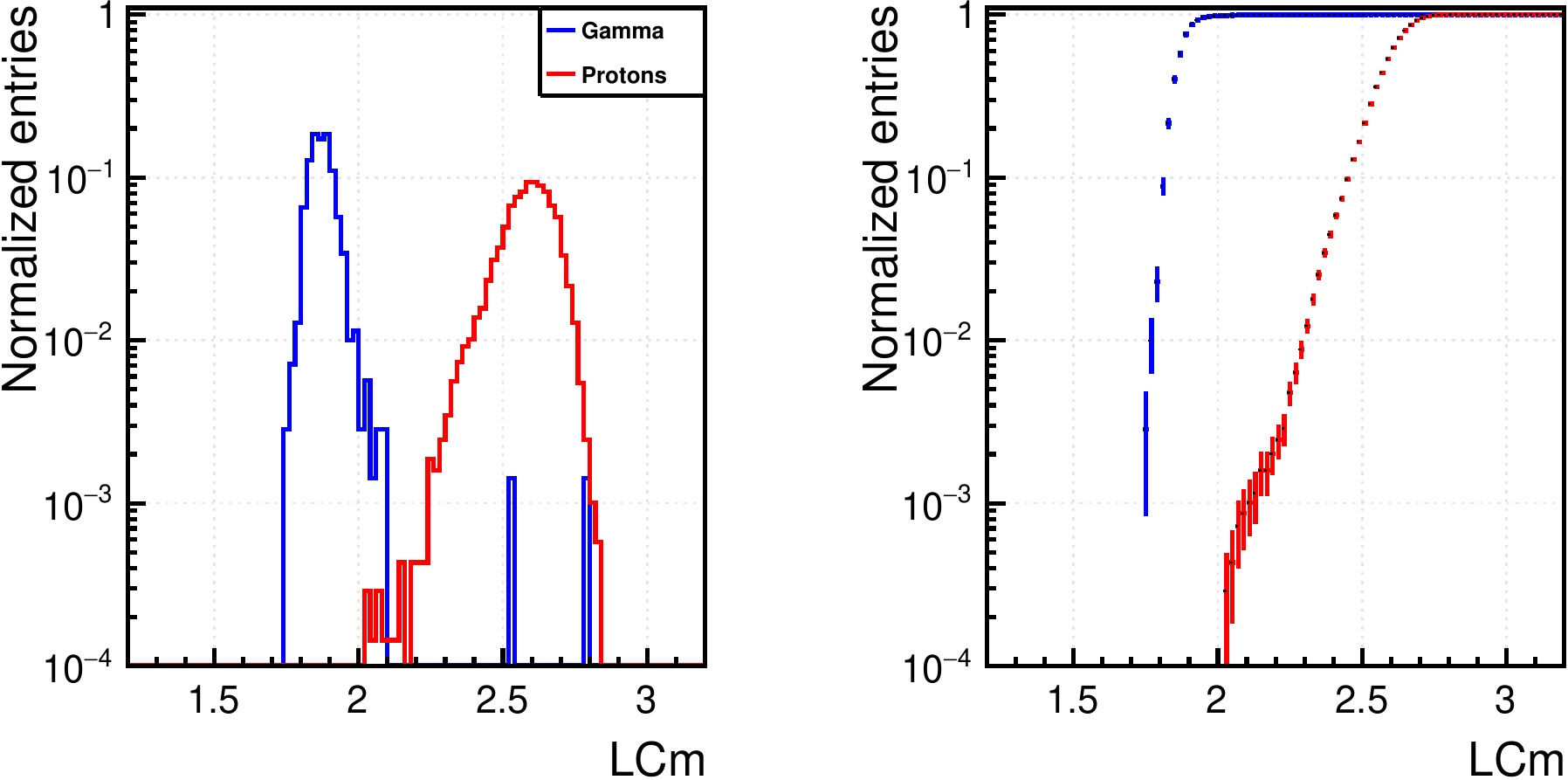}
\caption{\label{fig:LCm_100TeV_FF012}  $LCm$ distributions  for gamma showers with 100 TeV  (blue histogram, to the left) and for proton showers (red histogram, to the right) with  similar energies at ground considering an array with $FF=0.12$.
The sizes of the samples are $6745$ and $676$ events for protons and gammas, respectively.  On the left are the distributions, and on the right, the corresponding cumulatives.}
\end{figure}

The $LCm$ distributions as well as their cumulative distributions are  shown in figure \ref{fig:LCm_100TeV_FF012}  
for gamma showers with energies $\sim 100\,$TeV  (blue points) and for proton showers (red points) with similar energies at ground, considering an array with $FF=0.12$.
Within the currently limited statistics\footnote{$\mathcal{O}(10^4)$ events for proton showers.} no background events remain at a gamma efficiency close to $100\%$.

\begin{figure}[tbp]
 \centering
\includegraphics[width=0.8\linewidth]{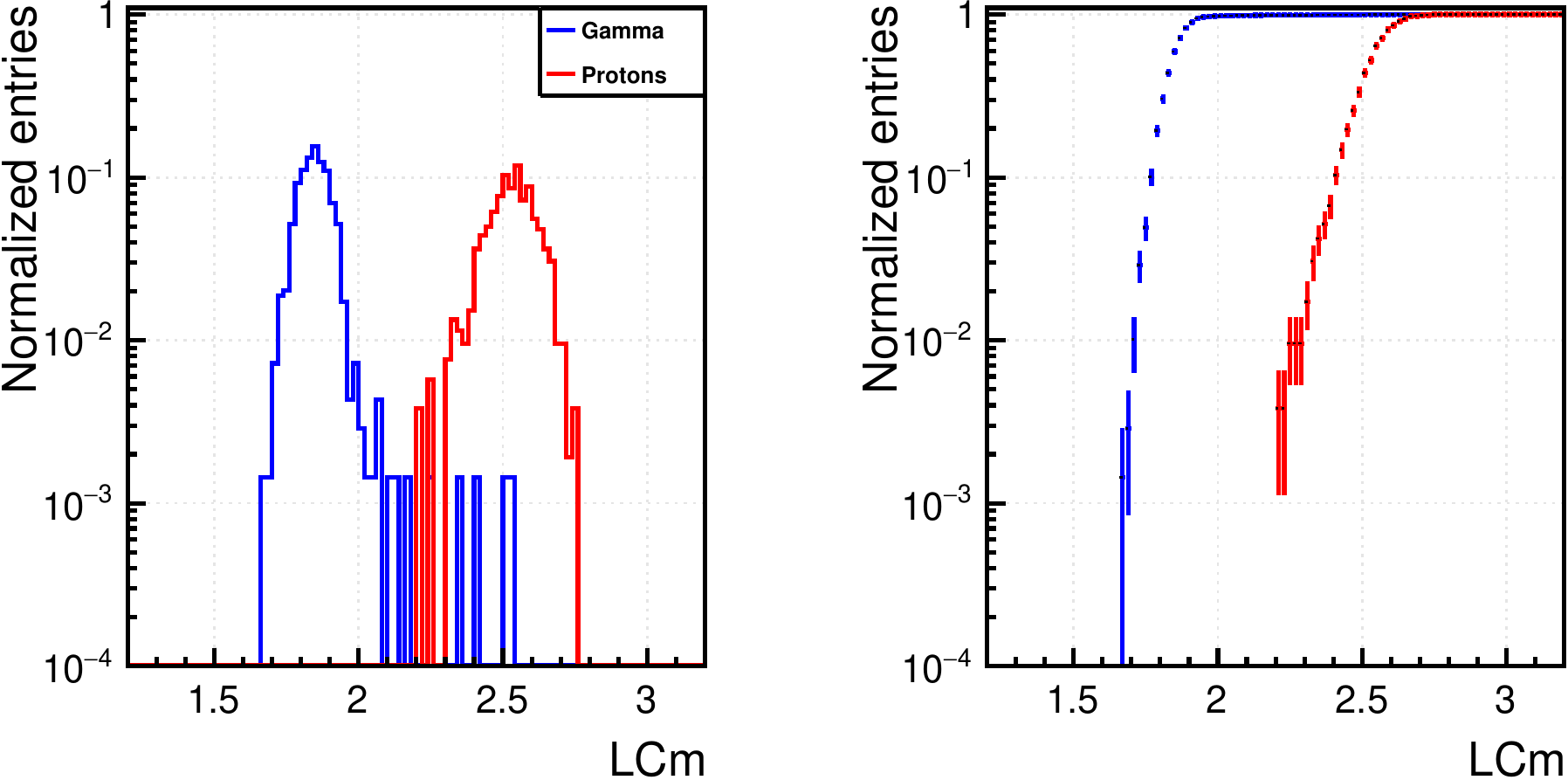}
\caption{\label{fig:LCm_1PeV_FF001} 
As  figure \ref{fig:LCm_100TeV_FF012} but considering  a $FF=0.014$ and the $1\,$PeV gamma energy bin.}
\end{figure}

\begin{figure}[tbp]
 \centering
\includegraphics[width=0.8\linewidth]{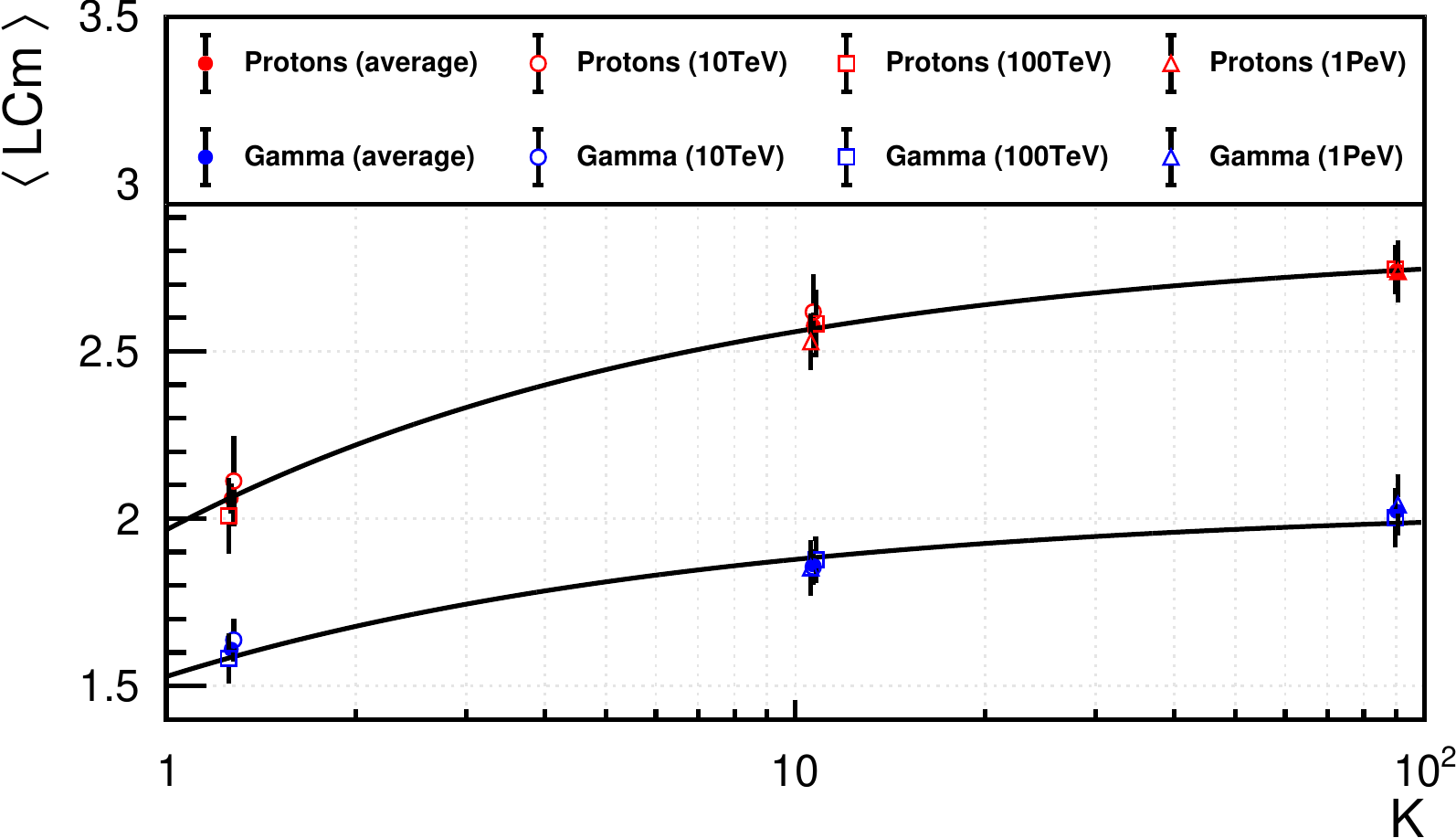}
\caption{\label{fig:LCm_K}  $\left< LCm \right>$ as a function of $K$ 
for gamma showers (blue, lower points) and for proton showers (red, upper points). The empty circles, squares and triangles correspond to samples with different energies,  as defined in  the inserted legend, and the full circles to the mean value for the several sets in  each K bin. The line is the best fit using the parametrisation defined by equation \ref{eq:LCm_K}.}
\end{figure}

The rule of thumb introduced in the previous section (equation \ref{eq:ruleThumb}) should also apply, by construction,  to the $LCm$ distributions, as it can be verified in figure \ref{fig:LCm_1PeV_FF001}, which is similar to figure \ref{fig:LCm_100TeV_FF012}, but now for the $1\,$PeV gamma energy bin and a $FF= 0.014$. In fact, $LCm$ is a function of $K$ which may be parameterised as (figure \ref{fig:LCm_K}):

\begin{equation}
\left< LCm_i \right> (K) \sim  A_i  + \frac{B_i}{\sqrt{K}} 
\label{eq:LCm_K}
\end{equation}

where $A_i$ and $B_i$ are constants defined for a given primary particle type ($i = \gamma$, proton). 
Asymptotically when $K \rightarrow \infty$ , $LCm_i(K) \rightarrow A_i $.
The values of $A_i$  were found to be $2.049$, $2.867$ for $\gamma$ and  proton, respectively, while the corresponding $B_i$ are equal to $-0.510$, $-0.874$.

\begin{figure}[tbp]
 \centering
\includegraphics[width=0.8\linewidth]{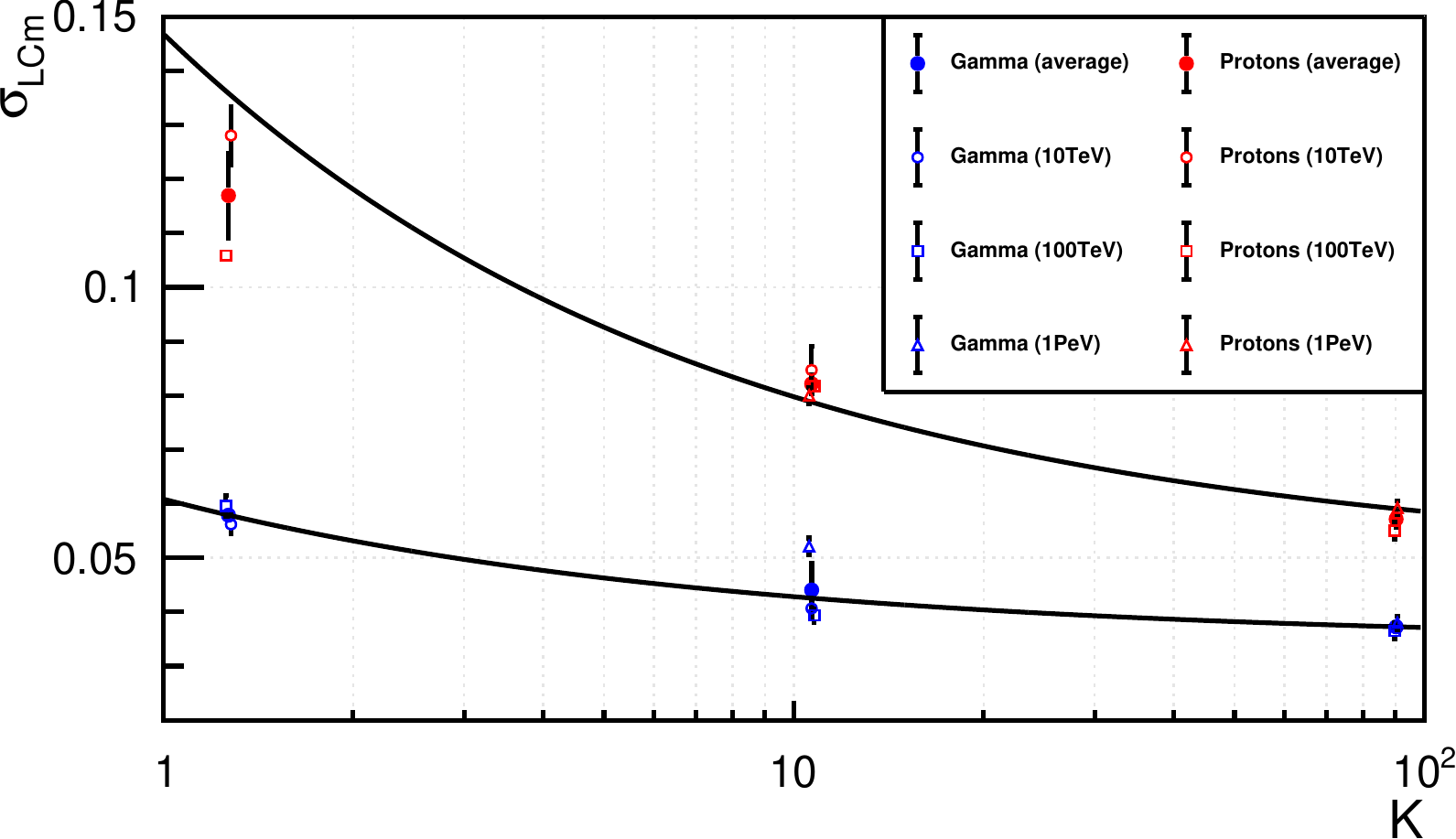}
\caption{\label{fig:LCm_width}  Width of the 
$LCm$ distributions as a function of $K$ for gamma showers (blue, lower points) and for proton showers (red, upper points). The empty circles, squares and triangles correspond to samples with different energies, as defined in the inserted legend, and the full circles to the mean value for the several sets in each K bin. The line is the best fit using the parametrisation defined by equation \ref{eq:LCM_witdth}.
}
\end{figure}

Furthermore, the width of the $LCm(r_{m})$ distributions for a given energy and fill factor was found to be reasonably well parametrized as function of $K$ (see figure \ref{fig:LCm_width}) using the following equation:

\begin{equation}
\sigma_{LCm_i} (K) \sim  \Delta_i + \frac{C_i}{\sqrt{K}}, 
\label{eq:LCM_witdth}
\end{equation}

where  $\Delta_i$ and $C_i$ are constants defined for each primary particle type.
Assuming Gaussian distributions, the values of $\Delta_i$  were found to be $0.036$, $0.055$ for $\gamma$ and proton, respectively, while the corresponding $C_i$
were $0.024$, $0.075$. 

These empirical parametrisations are thus well verified for all the simulated sets. However, to better establish the range of their validity and interpret their parameters in terms of the physics of the shower development, detailed studies implying larger simulated sets, different generators and realistic detector models will be needed.




\section{Correlation of $LCm$ with the total number of detected muons }
\label{sec:Ck-MU}

 The  number of detected muons is, as mentioned in section \ref{sec:introd}, often used as a $\gamma / {\rm h}$ discriminator. Hence, some degree of correlation between  $LCm$ and $N_{\mu}$ is expected. Muons are produced in the hadronic sub-showers, which are also the origin of the azimuthal non-uniformity. 

\begin{figure}[tbp]
 \centering
\includegraphics[width=0.8\linewidth]{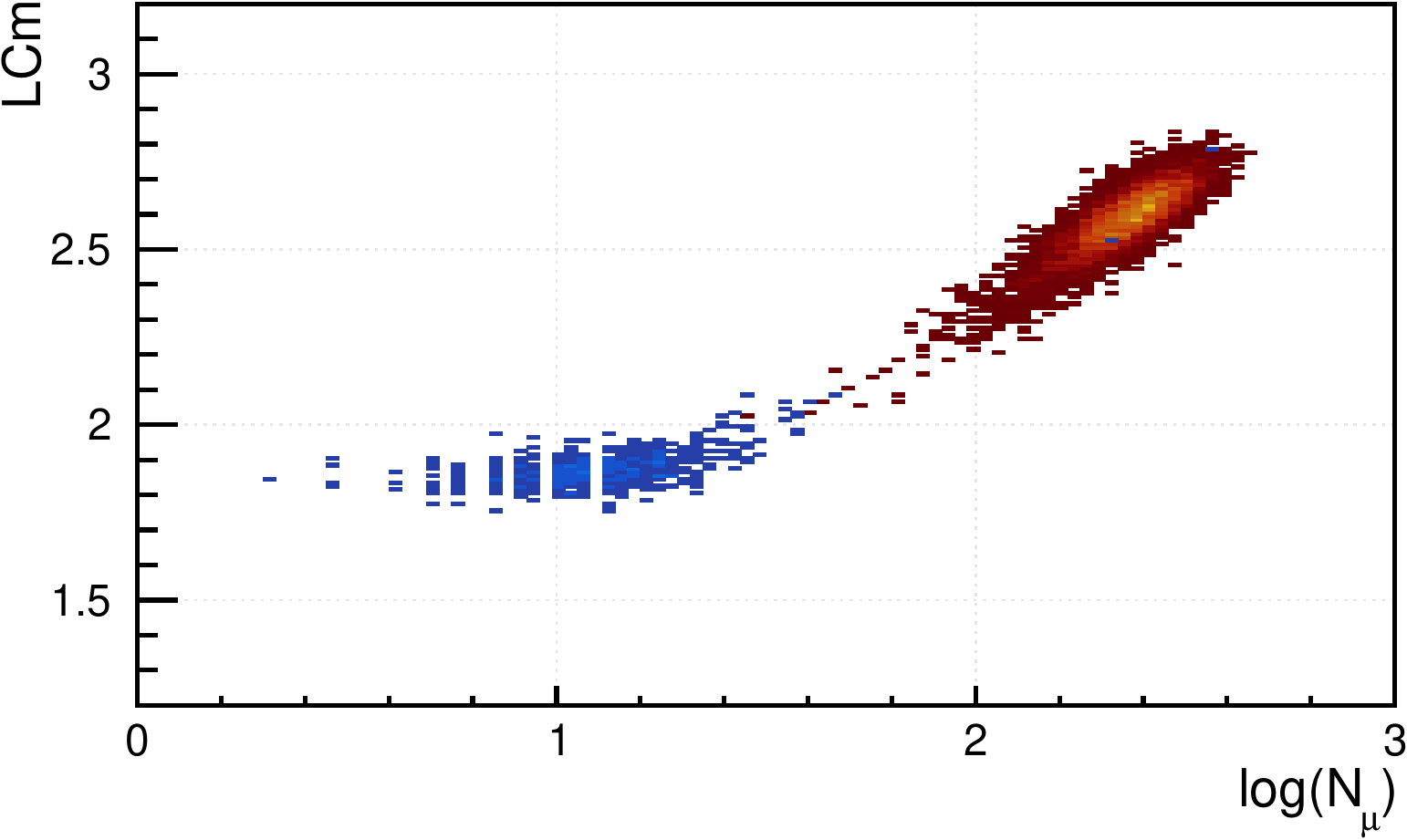}
\caption{\label{fig:LCm_Nmu_100_012_1} $LCm$ vs $\log(N_{\mu})$   distributions  for gamma showers with a primary energy of about 100 TeV (blueish histogram, to the left) and for proton showers (reddish histogram, to the right) with  similar energies at ground, considering an array with $FF=0.12$ and  a radius of 1000 m and placing the shower cores at the centre of the array. 
  }
\end{figure}

In figure \ref{fig:LCm_Nmu_100_012_1} we show, for the 100 TeV energy bin, the correlation  of the $LCm$ distribution with the  number of muons that hit the array stations located at a distance greater than $40\,$m from the shower core ($N_{\mu}$). The blue points  are the gamma showers and 
the red points are the proton-induced showers. A uniform array  with a fill factor of $12\%$ and a radius of $1000\,$m was considered. The  shower cores were placed at the centre of the array and the detection efficiency of the muons in the stations was considered to be 1. 
There is a clear almost linear correlation between $LCm$ and $\log(N_{\mu})$ for $N_{\mu} \gtrsim 15$.  It may be noted that one  gamma event has a high positive fluctuation on the detected number of muons with a similar positive fluctuation of the $LCm$ values. 
Therefore, within the present limited statistics, both $N_{\mu}$ and $LCm$ lead to equivalent background rejection factors for a high-efficiency gamma event tagging at an energy of 100 TeV.

\begin{figure}[tbp]
 \centering
\includegraphics[width=0.8\linewidth]{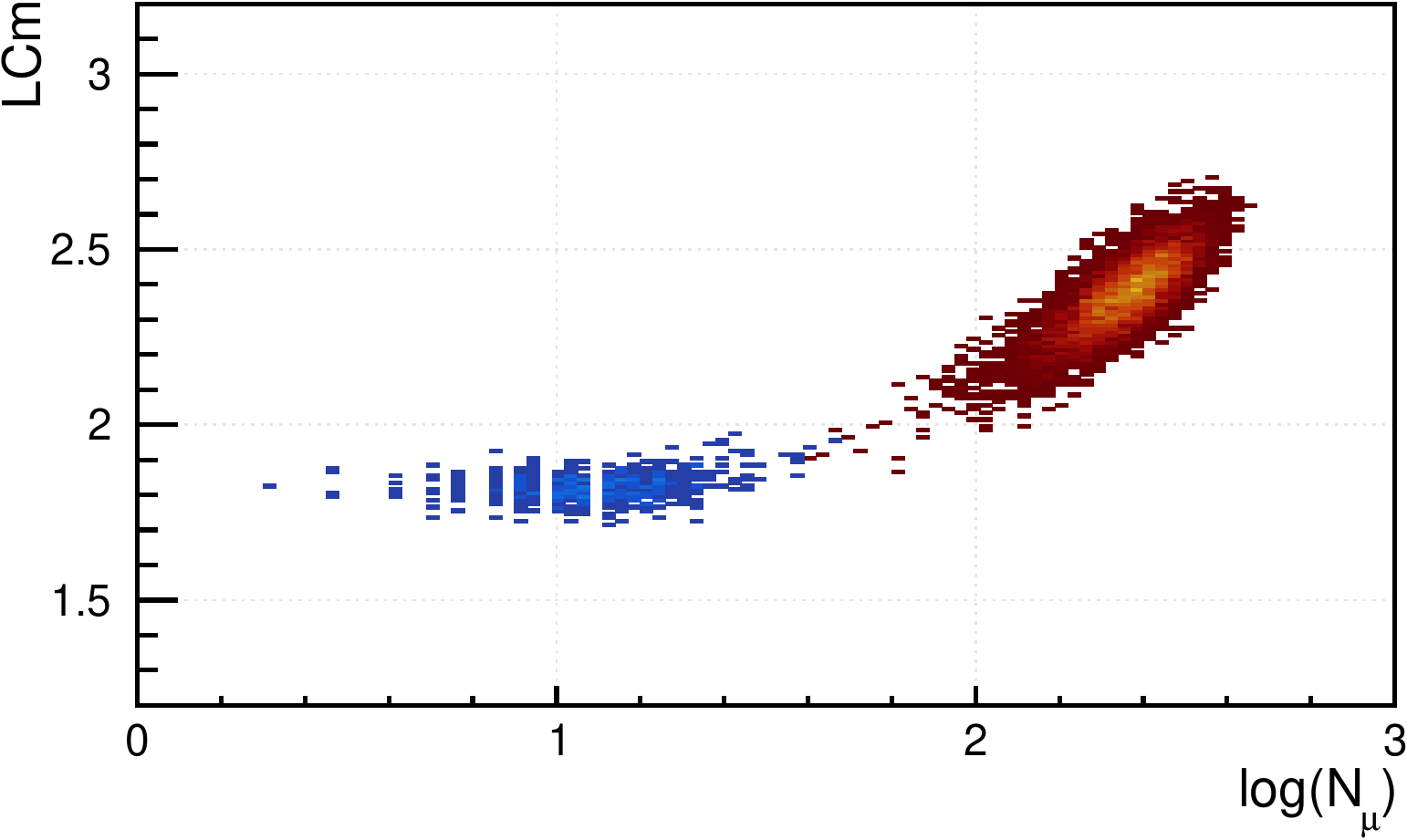}
\caption{\label{fig:LCm_Nmu_100_012_0} Same as figure \ref{fig:LCm_Nmu_100_012_1}, but not considering the contributions of muons and hadrons to the total signal of the stations.}
\end{figure}

The same distributions are reported in figure \ref{fig:LCm_Nmu_100_012_0} without considering now the contributions to the signal of the muons that hit the stations. The patterns shown in figures 
\ref{fig:LCm_Nmu_100_012_1} and \ref{fig:LCm_Nmu_100_012_0}
are basically identical. This non-trivial attribute suggests that the $LCm$ variable is indeed sensitive to the sub-clusters structure of the showers.
It is important to note that while these distributions are shown for an energy bin, the same correlation is present when analysing showers with fixed energy. This confirms that the correlation arises from intrinsic shower features rather than primary energy.

\section{Discussion and conclusions}
\label{sec:conclusions}

The recent identification in the Northern sky of more than 12 sources of gamma rays with energies up to the PeV opens a new era. These extreme energy events will be measured by present and future large wide-field gamma-ray observatories covering large surfaces and able to attain high background rejection factors. 
The measurement of the number of muons arriving at the ground is recognised, so far, as the best approach to reach such levels of rejection ($ > 10^4 $), but it leads to very costly detection solutions.

In this article, we show that the quantification of the azimuthal non-uniformity in the pattern of the shower at the ground as a function of the distance to the shower core may be an alternative way to access the intrinsic differences in the development of electromagnetic and hadronic showers without implementing any costly strategy to absorb the electromagnetic component of the shower.

A new quantity is introduced which is simple and easy to compute. This variable presents a discrimination power similar to the one attained by the detected number of muons at the ground. It is worth noticing that while very effective in rejecting hadronic-induced showers, the ultimate discrimination power should be reached by combining this new quantity with other available discriminators accessible in the experiment, particularly with the LDF based variables.

It should be noted that, while the simulations presented in this work were generated for vertical events, the discriminator $LCm$ has a similar discrimination capability for inclined showers, provided that $r_k$ is computed in the plane perpendicular to the shower axis. Such was verified using a small set of simulations with zenith angles up to $35^\circ$.

The study of the precise requirements for the design of future large ground array gamma-ray observatories, like the future Southern Wide-field Gamma-ray Observatory (SWGO) \cite{SWGO}, are out of the scope of this article. Such study would imply much larger sets of simulations both at the generator level and at the detector level to realistically include the performance of the chosen detector stations. However, we envisage that the station unit should be sensitive both to the muonic and to the electromagnetic component of the shower and it should have a low energy threshold. In this sense, water-Cherenkov detectors may be a good possibility. Moreover, the sparse array should cover in a way as uniform as possible surfaces of the order of $1\,{\rm km^2}$ or higher, with variable fill factor as a function of the distance to the centre of the array, ensuring  $K$ factors of about $10$ for each targeted energy in the relevant regions. For PeV energies, this implies fill factor values that may be lower than $1 \%$ in the outer regions of the sparse array. 

Finally, just a word to remark that this new proposed estimator may also contribute significantly to the study of charged cosmic rays and, namely, in the determination of the composition and in the test and validation of hadronic models in an extended energy range.

\acknowledgments

We would like to thank Jaime Alvarez-Mu\~niz, Pedro Brogueira, Roger Clay, Antonio Bueno, Alessandro De Angelis, Giorgio Matthiae, Jakub V\'icha, Alan Watson and Enrique Zas for carefully reading the manuscript and providing useful comments.
This work has been financed by national funds through FCT - Fundação para a Ciência e a Tecnologia, I.P., under project PTDC/FIS-PAR/4300/2020. R.~C.\ is grateful for the financial support by OE - Portugal, FCT, I. P., under DL57/2016/
cP1330/cT0002. 

\paragraph{Dedication} We dedicate this article to the memory of our friend, Prof. Ronald C. Shellard (Ron), who passed away recently, December 2021, and with whom some of us have worked closely during many years, developing new ideas and strategies, namely in the field of high energy charged and neutral cosmic rays.

\bibliographystyle{JHEP}
\bibliography{references.bib}

\providecommand{\href}[2]{#2}\begingroup\raggedright\begin{thebibliography}{10}

\bibitem{Peta_gamma}
M.~Amenomori et~al., \emph{{First Detection of sub-PeV Diffuse Gamma Rays from
  the Galactic Disk: Evidence for Ubiquitous Galactic Cosmic Rays beyond PeV
  Energies}}, \href{https://doi.org/10.1103/PhysRevLett.126.141101}{\emph{Phys.
  Rev. Lett.} {\bfseries 126} (2021) 141101}.

\bibitem{LHAASO_PeV}
{Cao, Zhen and et. al}, \emph{{Ultrahigh-energy photons up to 1.4
  petaelectronvolts from 12 $\gamma$ ray Galactic sources}},
  \href{https://doi.org/10.1038/s41586-021-03498-z}{\emph{\it Nature}
  {\bfseries 594} (2021) 33}.

\bibitem{Greisen}
K.~Greisen, \emph{Cosmic ray showers},
  \href{https://doi.org/10.1146/annurev.ns.10.120160.000431}{\emph{Annual
  Review of Nuclear Science} {\bfseries 10} (1960) 63}.

\bibitem{Abeysekara_2017}
Abeysekara and et. al, \emph{{Observation of the Crab Nebula with the HAWC
  Gamma-Ray Observatory}},
  \href{https://doi.org/10.3847/1538-4357/aa7555}{\emph{The Astrophysical
  Journal} {\bfseries 843} (2017) 39}.

\bibitem{LATTES}
P.~Assis et~al., \emph{{Design and expected performance of a novel hybrid
  detector for very-high-energy gamma-ray astrophysics}},
  \href{https://doi.org/10.1016/j.astropartphys.2018.02.004}{\emph{Astropart.
  Phys.} {\bfseries 99} (2018) 34}
  [\href{https://arxiv.org/abs/1607.03051}{{\ttfamily 1607.03051}}].

\bibitem{Hillas}
A.M.~Hillas, \emph{Cerenkov light images of eas produced by primary gamma},  in
  \emph{19th Intern. Cosmic Ray Conf-Vol. 3}, no.~OG-9.5-3, 1985.

\bibitem{Weekes}
T.C.~Weekes et~al., \emph{{Observation of TeV gamma rays from the Crab nebula
  using the atmospheric Cerenkov imaging technique}},
  \href{https://doi.org/10.1086/167599}{\emph{Astrophys. J.} {\bfseries 342}
  (1989) 379}.

\bibitem{Blake_1995}
P.R.~Blake and W.F.~Nash, \emph{Muons in extensive air showers. i. the lateral
  distribution of muons},
  \href{https://doi.org/10.1088/0954-3899/21/1/013}{\emph{Journal of Physics G:
  Nuclear and Particle Physics} {\bfseries 21} (1995) 129}.

\bibitem{Hayashida_1995}
N.~Hayashida and et.~al a, \emph{Muons ( $ \ge $ 1 {GeV}) in large extensive
  air showers of energies between 1016.5 ev and 1019.5 ev observed at akeno},
  \href{https://doi.org/10.1088/0954-3899/21/8/008}{\emph{Journal of Physics G:
  Nuclear and Particle Physics} {\bfseries 21} (1995) 1101}.

\bibitem{APEL2010202}
W.~Apel and et. al, \emph{The kascade-grande experiment},
  \href{https://doi.org/https://doi.org/10.1016/j.nima.2010.03.147}{\emph{Nuclear
  Instruments and Methods in Physics Research Section A: Accelerators,
  Spectrometers, Detectors and Associated Equipment} {\bfseries 620} (2010)
  202}.

\bibitem{1994NIMPA.346..329B}
A.~{Borione} and et. al, \emph{{A large air shower array to search for
  astrophysical sources emitting {\ensuremath{\gamma}}-rays with energies $\ge$
  10$^{14}$ eV}},
  \href{https://doi.org/10.1016/0168-9002(94)90722-6}{\emph{Nuclear Instruments
  and Methods in Physics Research A} {\bfseries 346} (1994) 329}.

\bibitem{Aab_2021}
A.~Aab and et. al, \emph{Calibration of the underground muon detector of the
  {P}ierre {A}uger {O}bservatory},
  \href{https://doi.org/10.1088/1748-0221/16/04/p04003}{\emph{Journal of
  Instrumentation} {\bfseries 16} (2021) P04003}.

\bibitem{LHAASO_muon}
{\scshape LHAASO} collaboration, \emph{{Design and performances of prototype
  muon detectors of LHAASO-KM2A}},
  \href{https://doi.org/10.1016/j.nima.2015.04.010}{\emph{Nucl. Instrum. Meth.
  A} {\bfseries 789} (2015) 143}.

\bibitem{MARTA}
P.~Abreu et~al., \emph{{MARTA: a high-energy cosmic-ray detector concept for
  high-accuracy muon measurement}},
  \href{https://doi.org/10.1140/epjc/s10052-018-5820-2}{\emph{Eur. Phys. J. C}
  {\bfseries 78} (2018) 333}
  [\href{https://arxiv.org/abs/1712.07685}{{\ttfamily 1712.07685}}].

\bibitem{DLWCD_Antoine}
A.~Letessier-Selvon, P.~Billoir, M.~Blanco, I.C.~Mari\c{s} and M.~Settimo,
  \emph{{Layered water Cherenkov detector for the study of ultra high energy
  cosmic rays}}, \href{https://doi.org/10.1016/j.nima.2014.08.029}{\emph{Nucl.
  Instrum. Meth. A} {\bfseries 767} (2014) 41}
  [\href{https://arxiv.org/abs/1405.5699}{{\ttfamily 1405.5699}}].

\bibitem{zuniga2017detection}
A.~Zu{\~n}iga-Reyes, A.~Hern{\'a}ndez, A.~Miranda-Aguilar, A.~Sandoval,
  J.~Mart{\'\i}nez-Castro, R.~Alfaro et~al., \emph{{Detection of vertical muons
  with the HAWC water Cherenkov detectors and its application to gamma/hadron
  discrimination}}, {\emph{arXiv preprint arXiv:1708.09500} (2017) }.

\bibitem{Gonzalez_2020}
B.S.~Gonz{\'{a}}lez and et. al, \emph{Using convolutional neural networks for
  muon detection in {WCD} tank},
  \href{https://doi.org/10.1088/1742-6596/1603/1/012024}{\emph{Journal of
  Physics: Conference Series} {\bfseries 1603} (2020) 012024}.

\bibitem{Mercedes}
P.~Assis et~al., \emph{{The Mercedes water Cherenkov detector}},
  \href{https://arxiv.org/abs/2203.08782}{{\ttfamily 2203.08782}}.

\bibitem{CORSIKA}
D.~Heck, J.~Knapp, J.~Capdevielle, G.~Schatz and T.~Thouw, \emph{A monte carlo
  code to simulate extensive air showers}, {\emph{Report FZKA} {\bfseries 6019}
  (1998) }.

\bibitem{fluka}
A.~Ferrari et~al., \emph{{FLUKA: A multi-particle transport code}},
  {\emph{CERN-2005-010, SLAC-R-773, INFN-TC-05-11} (2005) }.

\bibitem{fluka2}
T.~Böhlen et~al., \emph{{The FLUKA Code: Developments and Challenges for High
  Energy and Medical Applications}}, {\emph{Nuclear Data Sheets} {\bfseries
  120} (2014) 211}.

\bibitem{qgs}
S.~Ostapchenko, \emph{{Monte Carlo treatment of hadronic interactions in
  enhanced Pomeron scheme: I. QGSJET-II model}},
  \href{https://doi.org/10.1103/PhysRevD.83.014018}{\emph{Phys. Rev. D}
  {\bfseries 83} (2011) 014018}
  [\href{https://arxiv.org/abs/1010.1869}{{\ttfamily 1010.1869}}].

\bibitem{Borja4PMTs}
R.~Concei\c{c}\~ao, B.S.~Gonz\'alez, A.~Guill\'en, M.~Pimenta and B.~Tom\'e,
  \emph{{Muon identification in a compact single-layered water Cherenkov
  detector and gamma/hadron discrimination using machine learning techniques}},
  \href{https://doi.org/10.1140/epjc/s10052-021-09312-4}{\emph{Eur. Phys. J. C}
  {\bfseries 81} (2021) 542}
  [\href{https://arxiv.org/abs/2101.10109}{{\ttfamily 2101.10109}}].

\bibitem{CONEX}
T.~Bergmann et~al., \emph{One-dimensional hybrid approach to extensive air
  shower simulation}, {\emph{Astropart. Phys.} {\bfseries 26} (2007) 420}
  [\href{https://arxiv.org/abs/astro-ph/0606564}{{\ttfamily
  astro-ph/0606564}}].

\bibitem{SWGO}
P.~Huentemeyer et~al., \emph{{The Southern Wide-Field Gamma-Ray Observatory
  (SWGO): A Next-Generation Ground-Based Survey Instrument}}, {\emph{BAAS}
  {\bfseries 51} (2019) 109}.

\end{thebibliography}\endgroup

\end{document}